\documentclass[10pt] {article}
\usepackage[usenames,dvipsnames,svgnames,x11names]{xcolor}

\newcommand{\claim}[1]{\subsubsection{\textnormal{\textit{#1}}}}
\usepackage[T1]{fontenc} %
\usepackage[normalem]{ulem} %

\usepackage[nocompress]{cite}

\usepackage{listings}

\lstdefinelanguage{XML}
{
basicstyle=\ttfamily\footnotesize,
  morestring=[b]",
  moredelim=[s][\bfseries\color{Maroon}]{<}{\ },
  moredelim=[s][\bfseries\color{Maroon}]{</}{>},
  moredelim=[l][\bfseries\color{Maroon}]{/>},
  moredelim=[l][\bfseries\color{Maroon}]{>},
  morecomment=[s]{<?}{?>},
  morecomment=[s]{<!--}{-->},
  commentstyle=\color{gray},
  stringstyle=\color{blue},
  identifierstyle=\color{red}
}

\usepackage{moreverb}

\usepackage[nounderscore]{syntax}

\usepackage[pdftex]{graphicx}
\graphicspath{{./figures/}}
\DeclareGraphicsExtensions{.pdf}

\usepackage{subfig} %

\usepackage[cmex10]{amsmath}
\usepackage{amssymb}
\usepackage{mathtools}
\usepackage{amsthm}
\usepackage{amsfonts}

\usepackage{algorithmicx}
\usepackage{algpseudocode}
\usepackage[ruled]{algorithm}
\definecolor{light-gray}{gray}{0.75}
\algrenewcommand{\algorithmiccomment}[1]{\hskip3em{{\footnotesize \textcolor{light-gray}{$\blacktriangleright$}}} #1}

\usepackage{multirow} %
\usepackage{rotating} %
\usepackage{booktabs} %
\usepackage{colortbl} %
\usepackage{tablefootnote} %
\usepackage{array} %
\newcolumntype{L}[1]{>{\raggedright\let\newline\\\arraybackslash\hspace{0pt}}m{#1}}
\newcolumntype{C}[1]{>{\centering\let\newline\\\arraybackslash\hspace{0pt}}m{#1}}
\newcolumntype{R}[1]{>{\raggedleft\let\newline\\\arraybackslash\hspace{0pt}}m{#1}}

\usepackage[pdftex,colorlinks=true,urlcolor=blue,citecolor=blue]{hyperref}

\usepackage{xspace}

\usepackage{enumitem}

\newcommand{\mobilenet}{MobileNet\xspace}
\newcommand{\resnet}{ResNet\xspace}
\newcommand{\lenet}{LeNet\xspace}
\newcommand{\BM}{bare metal\xspace} %

\usepackage[english]{babel}

\usepackage[usenames,dvipsnames,svgnames,x11names]{xcolor}
\newcommand{\addc}[1]{{\textcolor{teal}{#1}}}
\newcommand{\delc}[1]{ {\textcolor{gray} {\sout{#1}} }}
\newcommand{\repc}[2]{ {\textcolor{gray} {\sout{#1}} }{\textcolor{teal} {#2}}}
\renewcommand{\addc}[1]{#1}
\renewcommand{\delc}[1]{}
\renewcommand{\repc}[2]{#2}

\begin{document}
\title{Performance Characterization of Containerized DNN Training and Inference on Edge Accelerators{\thanks{~\repc{Preprint}{Updated results} of short paper \repc{to appear}{published} in HiPC 2023: Prashanthi S.K., V. Hegde, K. Patchava, A. Das, and Y. Simmhan, “Performance Characterization of Containerized DNN Training and Inference on Edge Accelerators,” in \textit{IEEE International Conference on High Performance
Computing, Data, and Analytics (HiPC)}, 2023, \href{https://dx.doi.org/10.1109/HiPC58850.2023.00028}{doi:10.1109/HiPC58850.2023.00028}}}}

\author{Prashanthi S.K., Vinayaka Hegde, Keerthana Patchava,\\ Ankita Das and Yogesh Simmhan\\
Department of Computational and Data Sciences\\
Indian Institute of Science\\
Bangalore 560012 INDIA\\
Email: \{prashanthis, simmhan\} @iisc.ac.in
}
\date{}

\maketitle
\begin{abstract}
Edge devices have typically been used for DNN inferencing. The increase in the compute power of accelerated edges is leading to their use in DNN training also. As privacy becomes a concern on multi-tenant edge devices, Docker containers provide a lightweight virtualization mechanism to sandbox models. But their overheads for edge devices are not yet explored. In this work, we study the impact of containerized DNN inference and training workloads on an NVIDIA AGX Orin edge device and contrast it against bare metal execution on running time, CPU, GPU and memory utilization, and energy consumption. Our analysis \repc{provides several interesting insights on these overheads.}{shows that there are negligible containerization overheads for individually running DNN training and inference workloads.}
\end{abstract}

%\IEEEpeerreviewmaketitle

\section{Introduction}%

Edge devices such as the Raspberry Pi, Intel Movidius and Google Coral have typically been used to perform lightweight vision-based DNN inferencing in autonomous vehicles and smart city deployments. However, training on the edge is growing popular for two reasons. First is the advent of GPU-accelerated edge devices such as NVIDIA Jetsons that approach GPU workstations in compute power~\cite{prashanthi2023sigmetrics}. E.g., the recent AGX Orin features a 12-core ARM CPU, an Ampere GPU with $2048$ CUDA cores and $64$ tensor cores, and $32GB$ of shared RAM. This delivers 275~TOPS of compute that matches an RTX 3080 Ti desktop GPU, while being smaller than a paperback and operating within $60W$ of power. Second is the growing attention to data privacy and the rise of federated learning that trains DNN models on local data present on the edge, with
only model weights sent to the server~\cite{mcmahan_FL}.

As the capabilities of edge devices increase, they are deployed as infrastructure compute resources, e.g., in smart cities, with the provision for multi-tenancy. This requires them to sandbox the DNN application, be it training or inferencing, from the host for privacy and security, and eventually from other concurrent applications~\cite{continual_learning_iccv}.
Virtualization and containerization are two common techniques for such application packaging and sandboxing. 
Among these, Docker containers~\cite{docker} offer a lightweight means to balance isolation and efficiency.
While Docker is common in servers and workstations and its performance impact has been studied in detail, there are few investigations 
on the edge~\cite{IISWC_19, ICUAS18}, especially for training. 

The unique features of Jetson edge devices, such as  ARM-based CPUs, shared RAM between CPU and GPU, 
several power modes that control CPU, GPU and memory frequencies, and the lack of support for GPU partitioning across containers make them distinct from GPU servers and workstations.
This, coupled with their more modest relative performance, makes it necessary to thoroughly examine the overheads of containerized DNN training and inference on accelerated edges.

In this paper, we make the following specific contributions:
\begin{enumerate}
    \item We conduct detailed experiments on an NVIDIA Jetson Orin AGX device for $3$ representative DNN model--dataset combinations for training and inference using Docker containers and \BM.
    \item We characterize the overheads of containerization on the execution time, resource usage and energy consumption for DNN training and inference.
    \item We \addc{additionally} investigate \repc{and isolate the sources of the overheads}{containerization performance} for \delc{the DNN models with different computational characteristics by studying }various power modes on the device.
\end{enumerate}

Our insights can help users make informed design choices to configure edge devices and select appropriate workloads.

\subsection{Updates from the HiPC Version based on Sensitivity to Software Versions}
\addc{The results reported here are an updated version of those presented in our HiPC short paper~\cite{hipc}. We found that the PyTorch used by NVIDIA in L4T Containers for Jetson was based on a different commit of PyTorch 2.0.0 and not the release commit as indicated by the 2.0.0 tag (which was used in \BM). This difference in Pytorch commits across \BM and container caused a performance difference between bare metal and container. We have modified the Dockerfile and ensured that both bare metal and container use the same release commit of PyTorch 2.0.0, and the updated results show that there is no significant performance difference between bare metal and container for all the training and inference workloads that we study.}
%~\footnote{https://forums.developer.nvidia.com/t/recommended-pytorch-version-for-jetpack-5-1/273534}

\section{Background}%

\subsection{DNN training and inference pipeline}
A generic \emph{DNN training pipeline} has $3$ stages -- fetch, pre-process and compute. During the fetch phase, a mini-batch of data is fetched from disk to main memory. Then pre-processing happens on the CPU, where operations such as transformation or crop are done. Finally, the forward pass, backward pass and parameter update computation are done on the GPU. Training is run over all minibatches in an epoch, and several epochs are executed till convergence. In a typical \emph{inference pipeline}, data may arrive continuously over the network and be grouped into a fixed-size batch in-memory, before it is pre-processed on the CPU and the inference computation, i.e., the forward pass, happens on GPU. 

\subsection{Docker containers}
Docker~\cite{docker} is a popular containerization platform that provides a convenient method to package and deploy an application and its dependencies. Docker containers share the device's operating system kernel, which makes them more lightweight than virtual machines. Containers use Linux \textit{namespaces} to provide isolation between containers, and Linux \textit{cgroups} to partition and limit the access of containers to system resources. Containers are created from a Docker image, which is generated by executing initialization commands in a Dockerfile. Docker containers do not need hardware virtualization support.

\section{Related Work} %
\subsection{Virtualization studies on edge devices}
A few papers have examined virtualization on diverse edge devices. Roberto~\cite{morabito_edge} studies the performance of Docker on different Raspberry Pi and Odroid edge devices using various CPU, network, memory and disk benchmarks. Similarly, Hadidi et al.~\cite{IISWC_19} evaluate the performance of containerized DNN inferencing on several low-end edge devices like the Raspberry Pi. RPi-class devices are much more constrained than Jetson devices which have GPU accelerators, faster CPUs and larger memory. Some studies~\cite{ICUAS18} compare the performance of KVM and Docker on a Jetson TX2 using compute and network benchmarks. Divide and Save~\cite{dividesave_arxiv} presents a method to speed up computations by splitting up DNN inference workloads among multiple containers on the Jetson TX2 and AGX Orin. However, they only study CPU-based inference and omit the GPU accelerator, which offers much of the compute power. None of these consider DNN training on accelerated edges.

\subsection{Virtualization studies on servers and cloud}
 Xavier et al.~\cite{xavier_hpc} characterize the performance of various container and hypervisor-based virtualization techniques for HPC using compute, memory, network and disk benchmarks. Zhang et al.~\cite{hotcloud_19} characterize and predict the performance interference of GPU virtualization in cloud GPUs. It is interesting to note that GPU servers support various multi-tenancy and GPU partitioning mechanisms such as CUDA MPS (Multi Process Service) and MIG (Multi-Instance GPU). However, the Jetson class of edge devices does not support any of these mechanisms and time-shares the GPU among containers.

To the best of our knowledge, we are the first to perform a characterization of containerized DNN inference and training workloads on accelerated edge devices.

\section{Experiment Methodology} %
\subsection{Hardware platform}

We perform all our experiments on the Jetson AGX Orin developer kit~\cite{Orin}, NVIDIA's latest and most powerful accelerated edge device. The AGX Orin has $12$ ARM A78AE CPU cores, an Ampere GPU with $2048$ CUDA cores and $32$ GB of LPDDR5 RAM (no ECC) shared between the CPU and GPU. Its peak power is $60W$ and costs around $2000$ USD. It comes with the OS installed on eMMC (flash based storage) and supports a variety of other storage media such as Micro SD card, USB HDD and NVME SSD. The full specifications of the device can be found in Table~\ref{tbl:jetsonspecs}.
The AGX Orin offers a choice of several thousand custom power modes ($\approx$ $18k$), and each power mode can be thought of as a tuple of CPU cores, CPU frequency, GPU frequency and memory frequency. Unless otherwise mentioned, we run our experiments in the MAXN power mode shown in Table~\ref{tbl:power}, where all components are set to their maximum possible frequencies. Dynamic Voltage and Frequency Scaling (DVFS) is off, and the onboard fan is set to maximum speed to avoid any temperature throttling effects. We use a $250$GB NVME Samsung SSD 980 with a sequential read speed of about $3.5$GBps to store the datasets, and this is exposed to containers as a Docker volume.
\begin{table}[t]
\centering
\footnotesize
% \vspace{-0.1in}
\caption{Specifications of NVIDIA Jetson AGX Orin devkit}
\label{tbl:jetsonspecs}
% \vspace{-0.1in}
\begin{tabular}{L{3.5cm}|R{3.4cm}}
\toprule
\textbf{Feature} & {AGX Orin} \\
\midrule
CPU Architecture & ARM Cortex A78AE\\ 
\noalign{\global\arrayrulewidth=0.1pt}\arrayrulecolor{lightgray}\hline
\noalign{\global\arrayrulewidth=0.4pt}\arrayrulecolor{black}
CPU Cores$^\dagger$ & 12\\
\noalign{\global\arrayrulewidth=0.1pt}\arrayrulecolor{lightgray}\hline
\noalign{\global\arrayrulewidth=0.4pt}\arrayrulecolor{black}
CPU Frequency (MHz)$^\dagger$ & 2200\\ \hline
GPU Architecture & Ampere\\
\noalign{\global\arrayrulewidth=0.1pt}\arrayrulecolor{lightgray}\hline
\noalign{\global\arrayrulewidth=0.4pt}\arrayrulecolor{black}
CUDA/Tensor Cores & 2048/64\\
\noalign{\global\arrayrulewidth=0.1pt}\arrayrulecolor{lightgray}\hline
\noalign{\global\arrayrulewidth=0.4pt}\arrayrulecolor{black}
GPU Frequency (MHz)$^\dagger$ & 1300\\ \hline
RAM (GB) & 32, LPDDR5\\ \hline
Storage Interfaces & $\mu$SD, eMMC, NVMe, USB \\
\noalign{\global\arrayrulewidth=0.1pt}\arrayrulecolor{lightgray}\hline
\noalign{\global\arrayrulewidth=0.4pt}\arrayrulecolor{black}
Memory Frequency (MHz)$^\dagger$ & 3200\\ \hline
Peak Power (W) & 60\\ \hline
Price (USD) & $\$1999$  \\ \hline
Form factor (mm) & $110 \times 110 \times 71.65$ \\
\bottomrule
\multicolumn{2}{L{8cm}}{$^\dagger$~This is the maximum possible value across all power modes.
Actual value depends on the power mode used}
\end{tabular}
%\vspace{-0.15in}
\end{table}

\subsection{Software libraries}
The AGX Orin runs JetPack version $5.1$, which comes with Linux for Tegra (L4T) $r35.2.1$, CUDA $v11.4.315$ and cuDNN $v8.6.0.166$. We use PyTorch $v2.0$ as the DL framework with torchvision $v0.15.1$. We set the \textit{num\_workers} to $4$ in the PyTorch Dataloader to enable pipelined and parallel data fetch and pre-processing. Docker version $20.10.21$ is used for containerization. We use the NVIDIA L4T PyTorch container for Jetpack as our base container image. The container image runs the same JetPack, CUDA and cuDNN versions as the \BM. We also ensure that other libraries have consistent versions across \BM and containers. %

\begin{table*}[t]
\centering
\setlength{\tabcolsep}{1pt}
\footnotesize
% \vspace{-0.1in}
\caption{DNN Models and Datasets}
\label{tbl:modeldataset}
% \vspace{-0.1in}
\begin{tabular}
{|rR{1.2cm}R{1.2cm}R{1.2cm}R{1.6cm}|R{1.4cm}R{1.4cm}R{1cm}R{1.4cm}|c}
\toprule
 \bf{Model} & \bf{\# Layers} & \bf{\# Params} & \bf{FLOPS}$^\dagger$ & \bf{Dataset} & \bf{\# Samples} & \bf{Size on Disk} & \bf{Minibatch Size} & \bf{Used for}\\
  \midrule
 \textbf{\lenet -5} & 7 \cite{lenet} &  $60k$ &  $4.4M$   &  \textbf{MNIST} &  $60,000$   & $46MB$ &  $16$  &  Train, Inf\\
  \hline
 \textbf{{\mobilenet}\,v3}& 20 \cite{mobilenet} & $5.48M$ & $225.4M$ & \textbf{GLD23k} & $23,080$  & $2.82GB$ 
 & $16$ &  Train, Inf\\
  \hline
 \textbf{\resnet -18$^*$}& 18 \cite{resnet} & $11.68M$  & $1.82G$   
 & \textbf{ImageNet} & $50,000$ & $6.74GB$ & $16$  & Train \\
   \hline
 \textbf{\resnet -50}& 50 \cite{resnet} & $26M$ & $4G$   
 & \textbf{ImageNet} & $50,000$ & $6.74GB$ & $16$  & Inf \\
 
\bottomrule
\multicolumn{9}{L{11.8cm}}{$^\dagger$~ As per the typical practice, FLOPS reported corresponds to a forward pass with minibatch size 1. 
\quad
$^*$ We use a smaller \resnet for training to avoid running out of memory}
\end{tabular}
%\vspace{-0.15in}
\end{table*}

\subsection{Models and datasets} 
We choose three popular computer vision DNN models which perform image classification tasks for our experiments. These models provide different computational intensities and architectures and are representative of edge workloads. \lenet~\cite{lenet} is a simple Convolutional Neural Network designed to recognize handwritten digits, and we use it with the MNIST dataset. \mobilenet~\cite{mobilenet} is a lightweight model designed for mobile devices using hardware-aware Network Architecture Search techniques, and we use the MobileNetv3large variant with a subset of the Google Landmarks Dataset (GLD23k). We also use two Residual Neural Networks~\cite{resnet} - a smaller variant \resnet -18 for training and a larger variant \resnet -50 for inference, both with a subset of the ImageNet dataset. The details of the models and dataset are listed in Table~\ref{tbl:modeldataset}.
 
\subsection{Performance metrics} We measure and report several system parameters such as CPU and GPU utilization, RAM usage and power (sampled every $1s$). We use the \textit{jtop} Python module (a wrapper around the \textit{tegrastats} utility from NVIDIA) to measure CPU and GPU utilization and power. Power measurements are based on the onboard sensors that capture the module load, and these readings are aggregated over the runtime of the workload to capture total energy. The Linux utility \textit{free} is used to measure used, free and cached memory. We instrument the PyTorch code to measure fetch stall, GPU compute and end-to-end times for every training and inference minibatch. Fetch stall time is the time taken for fetching and pre-processing data that does not overlap with GPU compute, and results in the GPU being idle. GPU compute time is measured using \texttt{torch.cuda.event} with \texttt{synchronize} to accurately capture execution time on the GPU. We also measure the end-to-end execution time of the minibatch and report our logging overhead as the difference between end-to-end time and the sum of fetch stall and GPU compute times. We have performed all experiments multiple times to ensure reproducibility.

\section{Results and Analysis} %

\begin{figure}[t]
 % \vspace{-0.15in}
\centering%
  \subfloat[Average minibatch time]
  {
    \includegraphics[width=0.45\columnwidth]{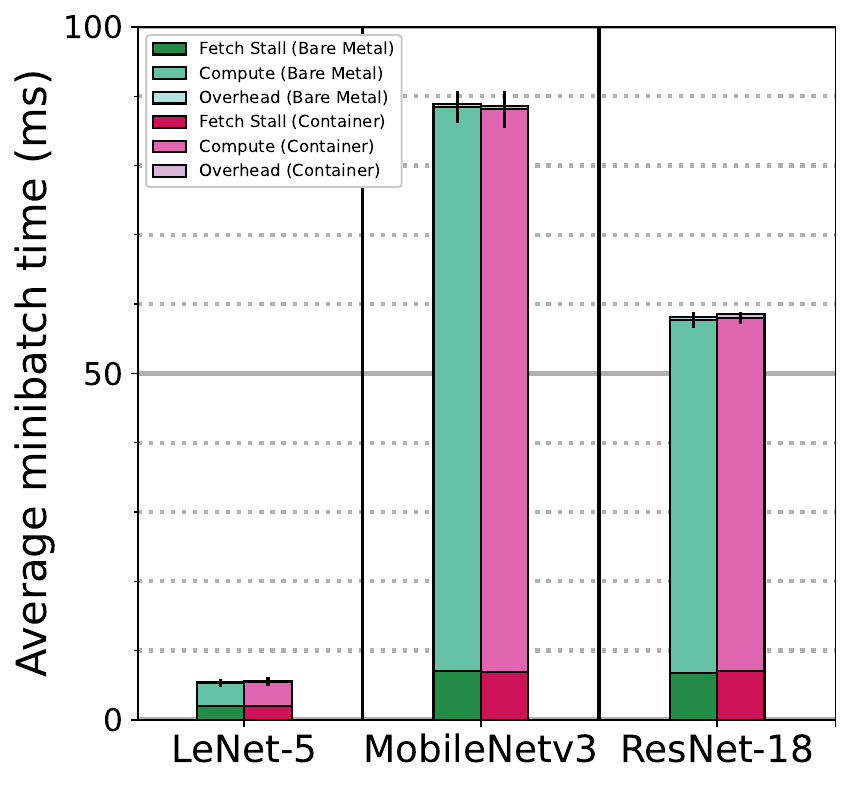}
    \label{fig:train:mbtime}
  }\quad
  \subfloat[Total energy]{
  \includegraphics[width=0.45\columnwidth]{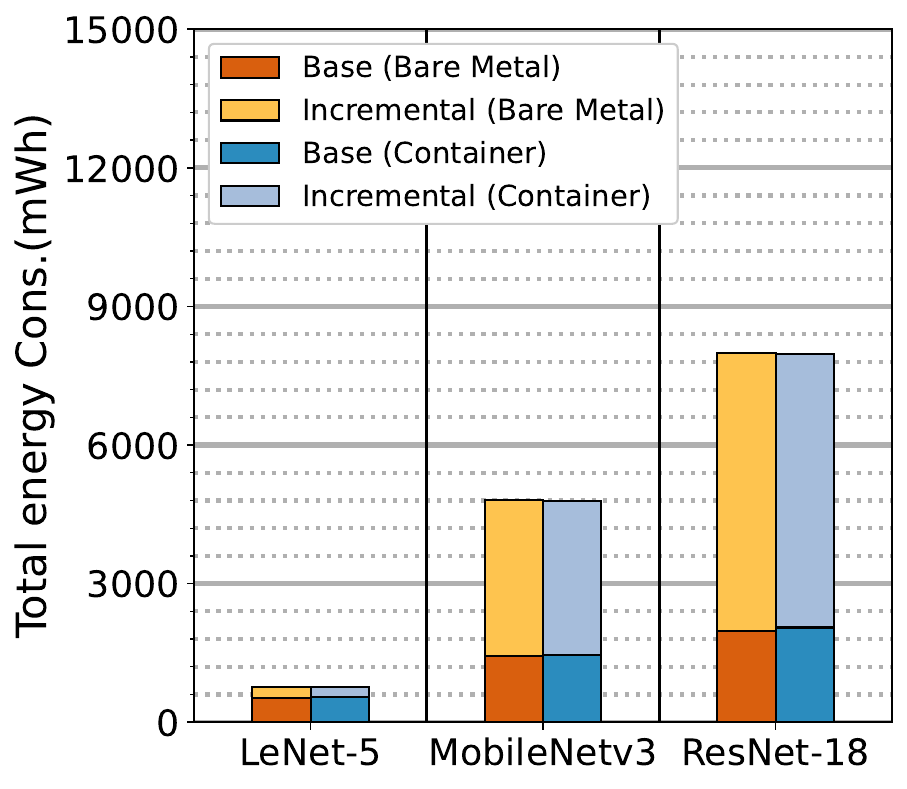}
    \label{fig:train:energy}
  }

\caption{Avg. minibatch time \& total energy for DNN training}%
\label{train_time_energy}
 % \vspace{-0.125in}
\end{figure}
\begin{figure}[t]
 % \vspace{-0.15in}
\centering%
  \subfloat[CPU utilization]
  {
    \includegraphics[width=0.32\columnwidth]{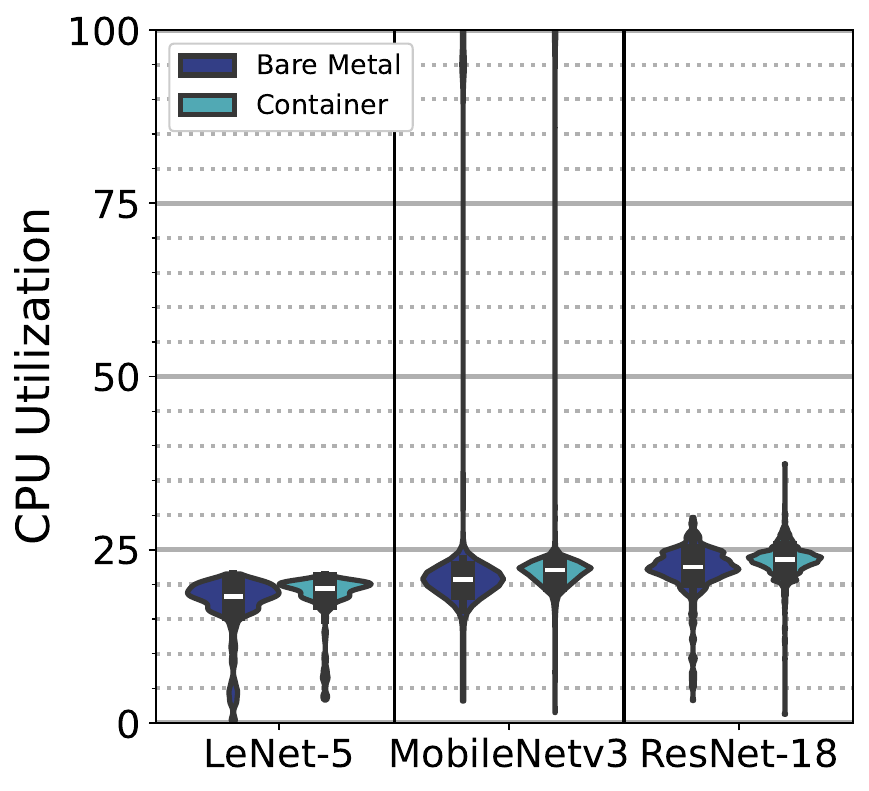}
    \label{fig:train:cpuutil}
  }
  \subfloat[GPU utilization]{
  \includegraphics[width=0.32\columnwidth]{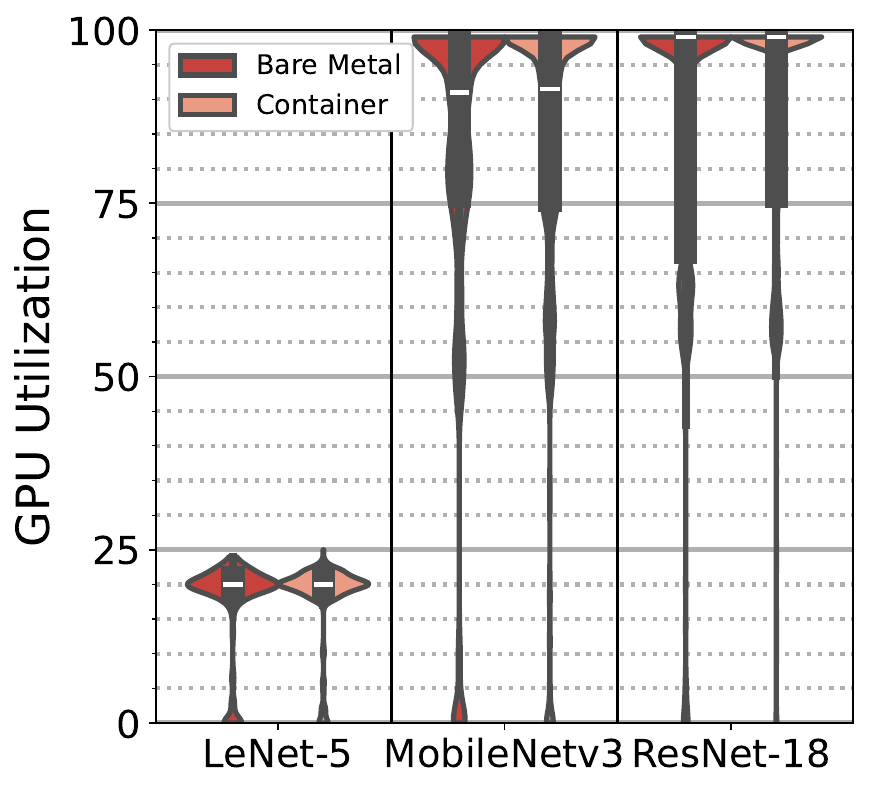}
    \label{fig:train:gpuutil}
  }
  \subfloat[Power]{
  \includegraphics[width=0.32\columnwidth]{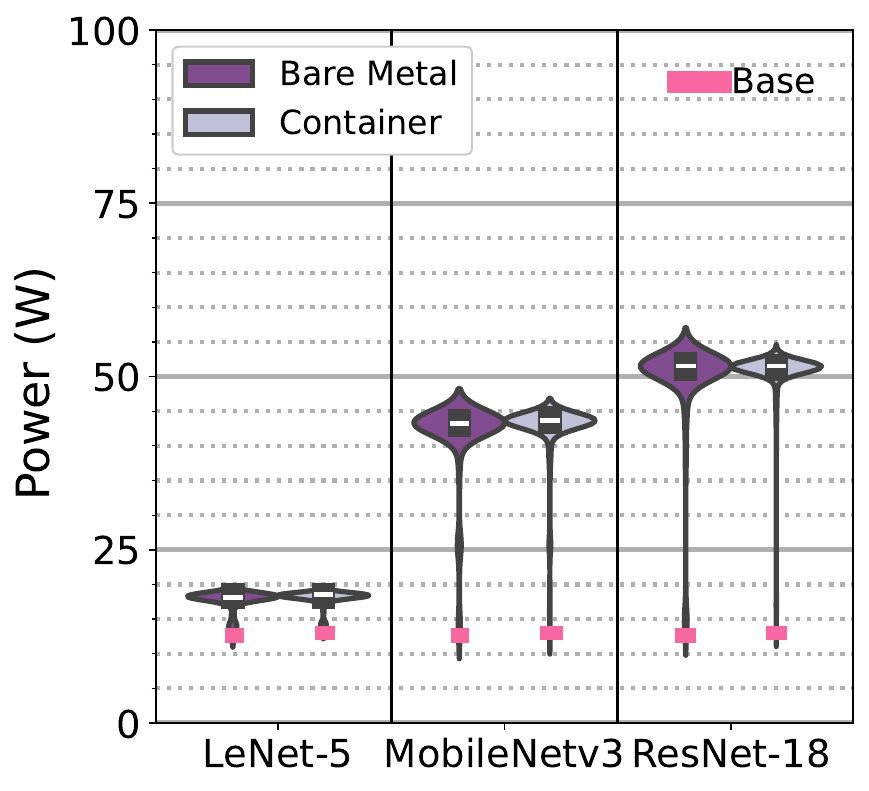}
    \label{fig:train:power}
  }
\caption{CPU, GPU utilization and power for DNN training}
\label{train_util_power}
 %\vspace{-0.25in}
\end{figure}
\begin{figure}[t]
 % \vspace{-0.15in}
\centering%
  \subfloat[Average minibatch time]
  {
    \includegraphics[width=0.45\columnwidth]{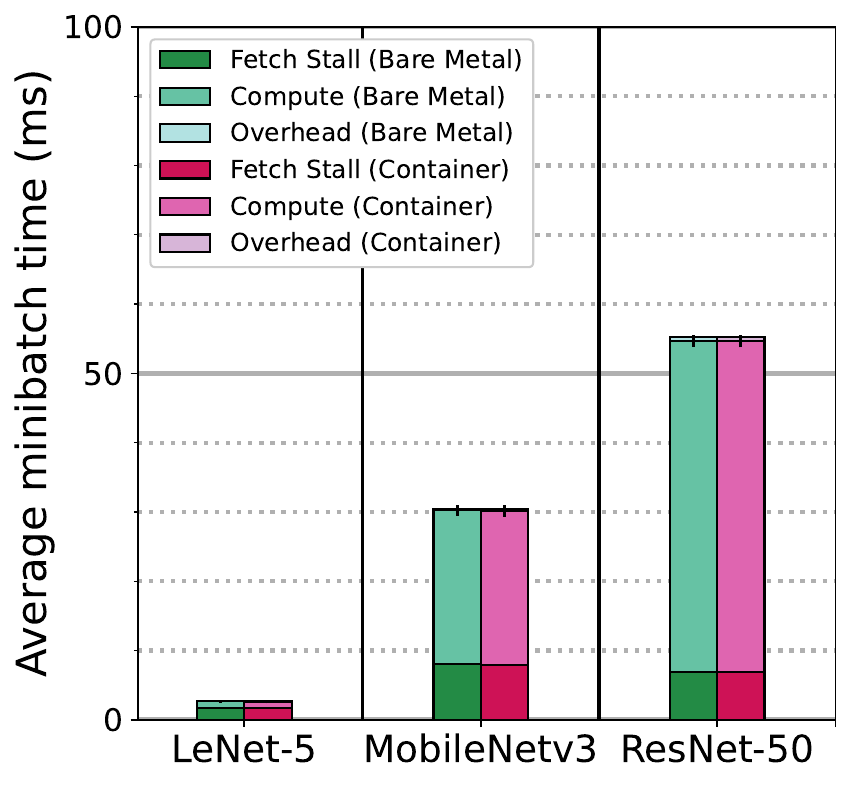}
    \label{fig:infer:mbtime}
  }\quad
  \subfloat[Total energy]{
  \includegraphics[width=0.45\columnwidth]{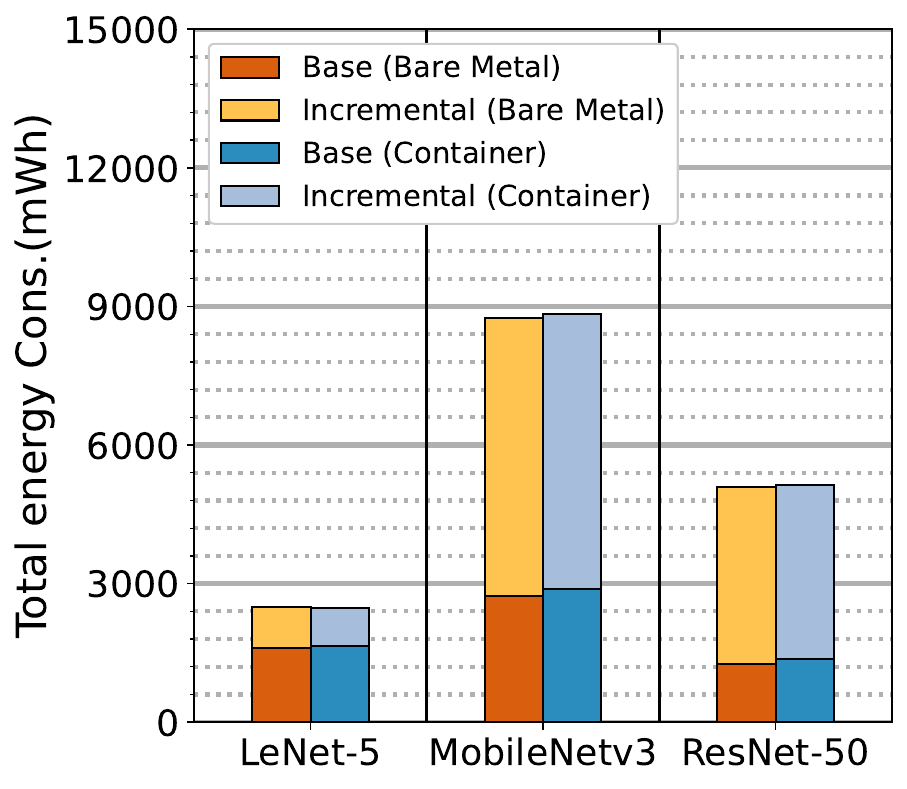}
    \label{fig:infer:energy}
  }
\caption{Avg. minibatch time \& total energy for DNN inf.} %
\label{infer_time_energy}
 % \vspace{-0.125in}
\end{figure}
\begin{figure}[t]
 %\vspace{-0.15in}
\centering%
  \subfloat[CPU utilization]
  {
    \includegraphics[width=0.32\columnwidth]{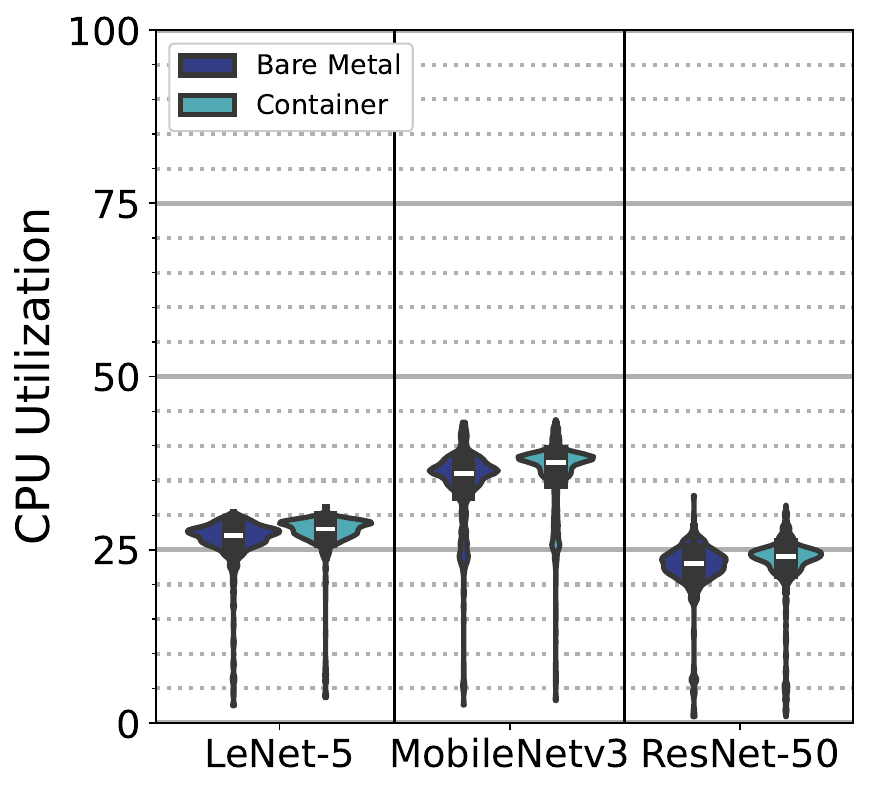}
    \label{fig:inf:cpuutil}
  }
  \subfloat[GPU utilization]{
  \includegraphics[width=0.32\columnwidth]{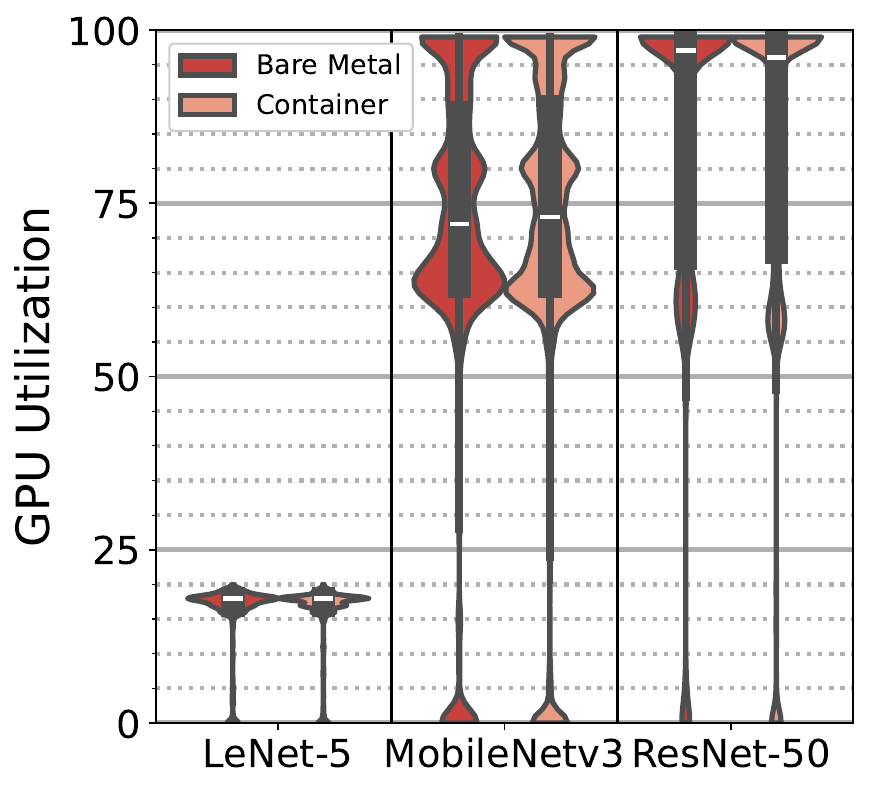}
    \label{fig:inf:gpuutil}
  }
  \subfloat[Power]{
  \includegraphics[width=0.32\columnwidth]{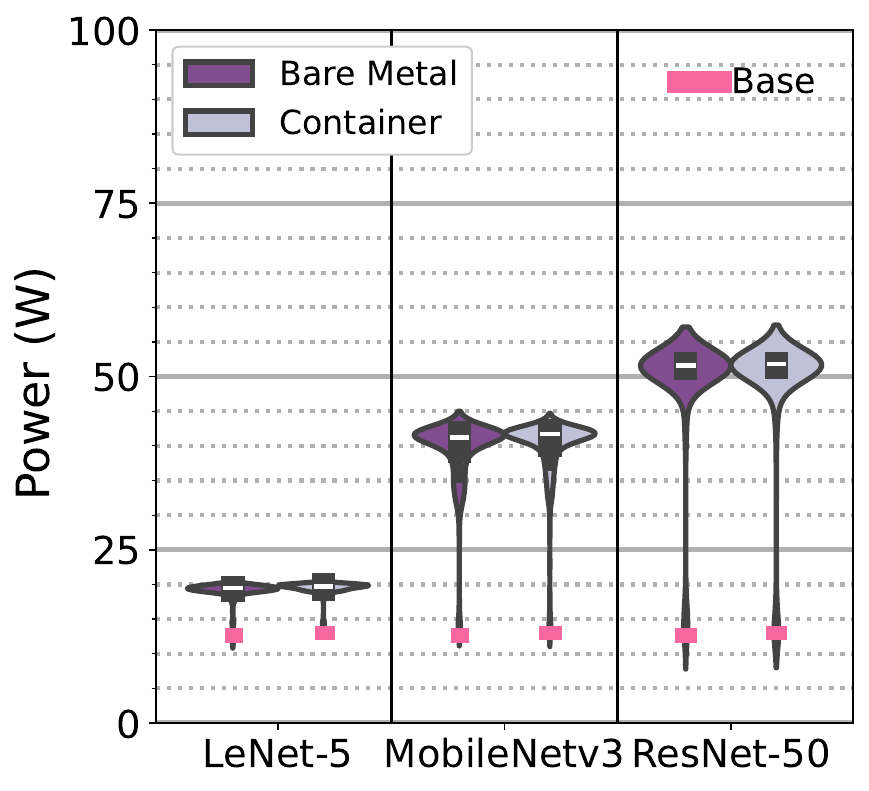}
    \label{fig:inf:power}
  }
\caption{CPU, GPU utilization and power for DNN inference}%
\label{inf_util_power}
%\vspace{-0.25in}
\end{figure}
\begin{figure}[t]
 % \vspace{-0.15in}
\centering%
  \subfloat[Training- memory footprint]
  {
    \includegraphics[width=0.35\columnwidth]{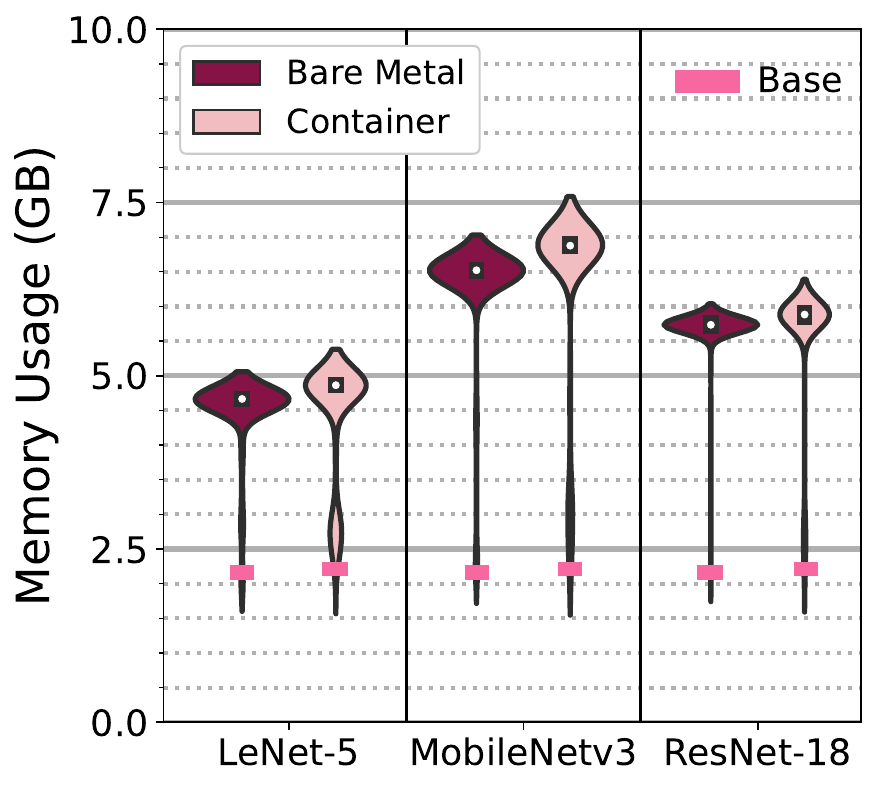}
    \label{fig:train:mem}
  }\quad
  \subfloat[Inference- memory footprint]{
  \includegraphics[width=0.35\columnwidth]{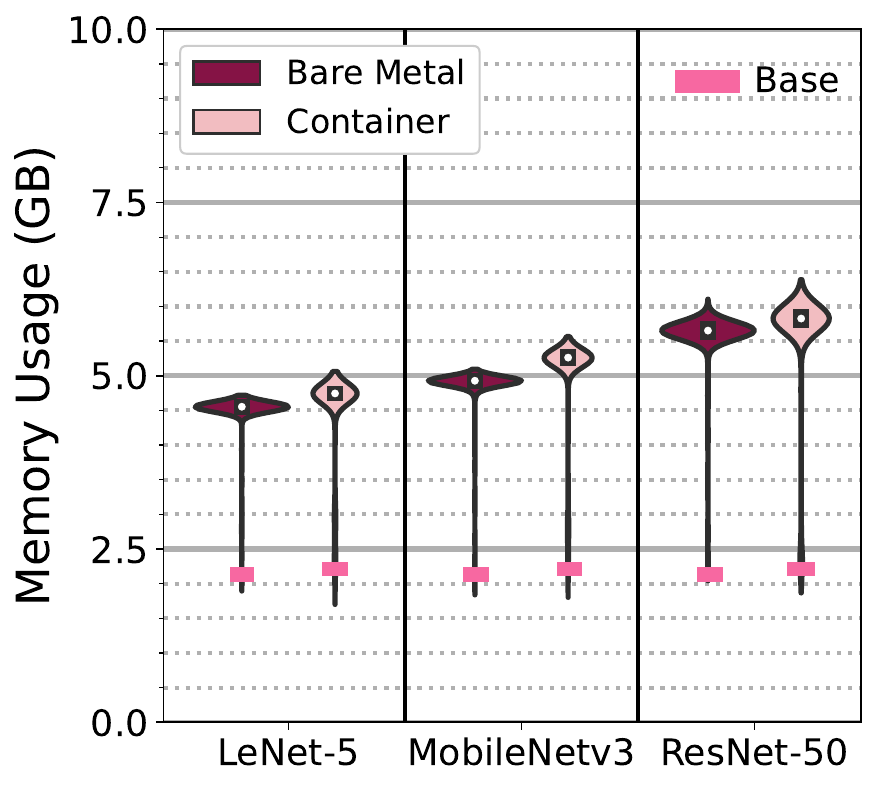}
    \label{fig:infer:mem}
  }
\caption{Memory footprint for DNN training and inference}
\label{train_infer_memory}
% \vspace{-0.25in}
\end{figure}

\begin{figure*}[t!]
 % \vspace{-0.15in}
\centering%
  \subfloat[Training - \lenet-5]
  {
    \includegraphics[width=0.45\columnwidth]{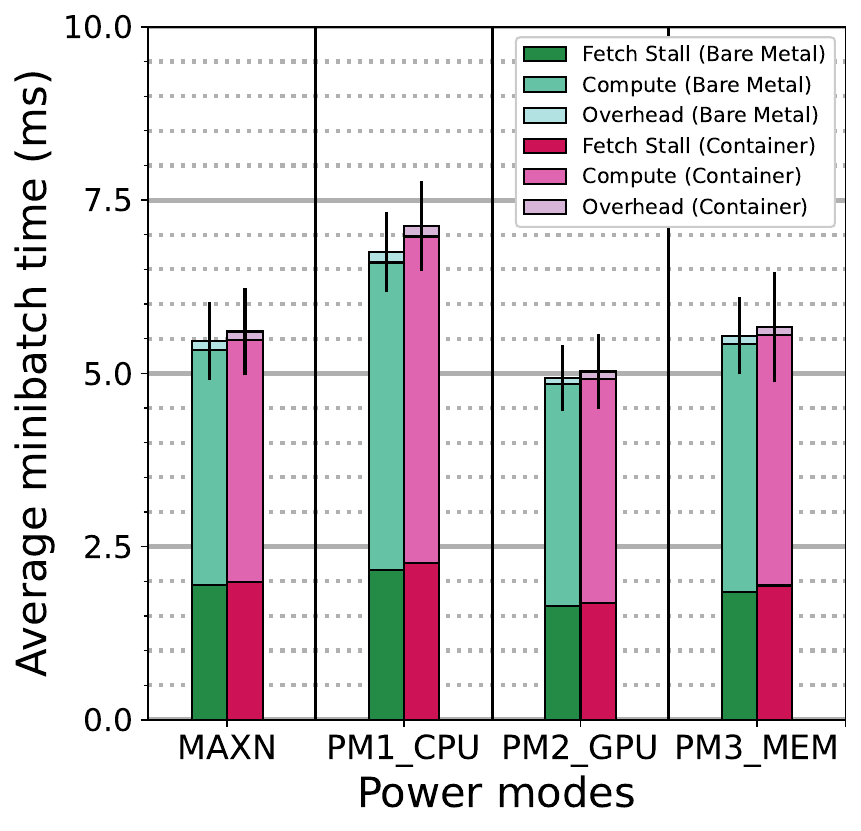}
    \label{fig:train:powermodes:lenet}
  }\quad
  \subfloat[Training - \mobilenet v3]{
  \includegraphics[width=0.45\columnwidth]{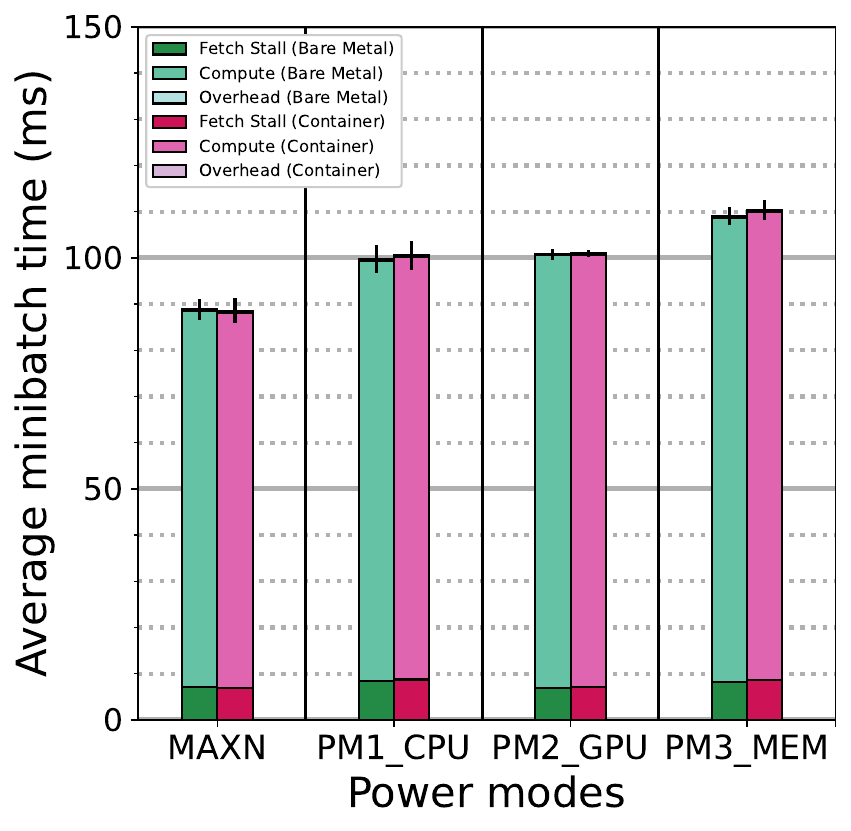}
    \label{fig:train:powermodes:mobilenet}
  }\quad
  \subfloat[Training - \resnet-18 ]{
  \includegraphics[width=0.45\columnwidth]{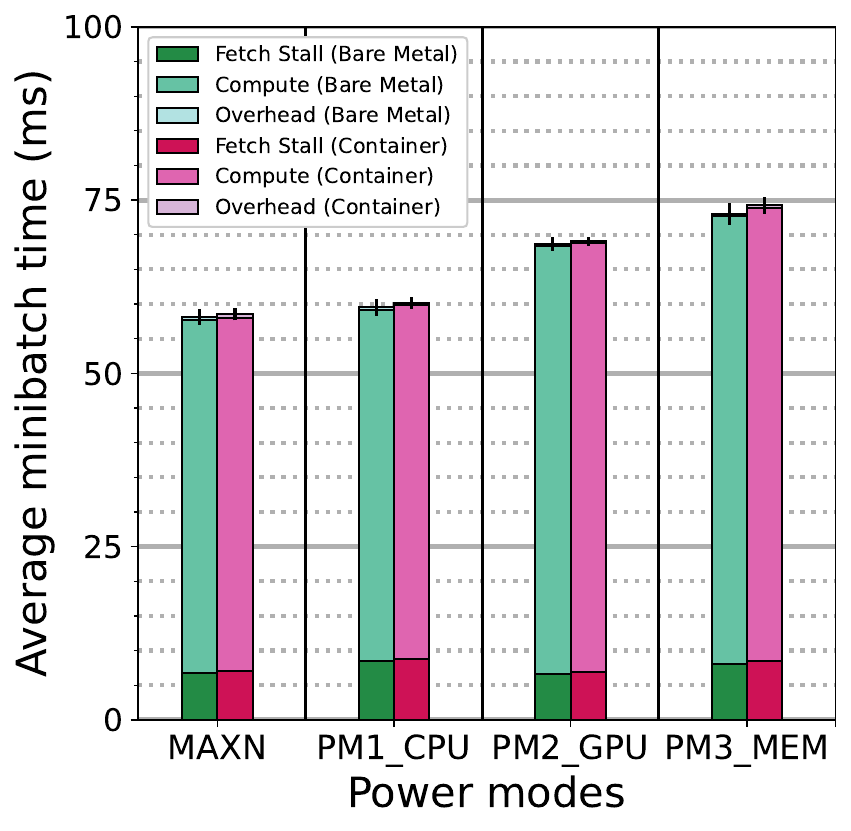}
    \label{fig:train:powermodes:resnet}
  }\quad
    \subfloat[Inference - \mobilenet v3]{
  \includegraphics[width=0.45\columnwidth]{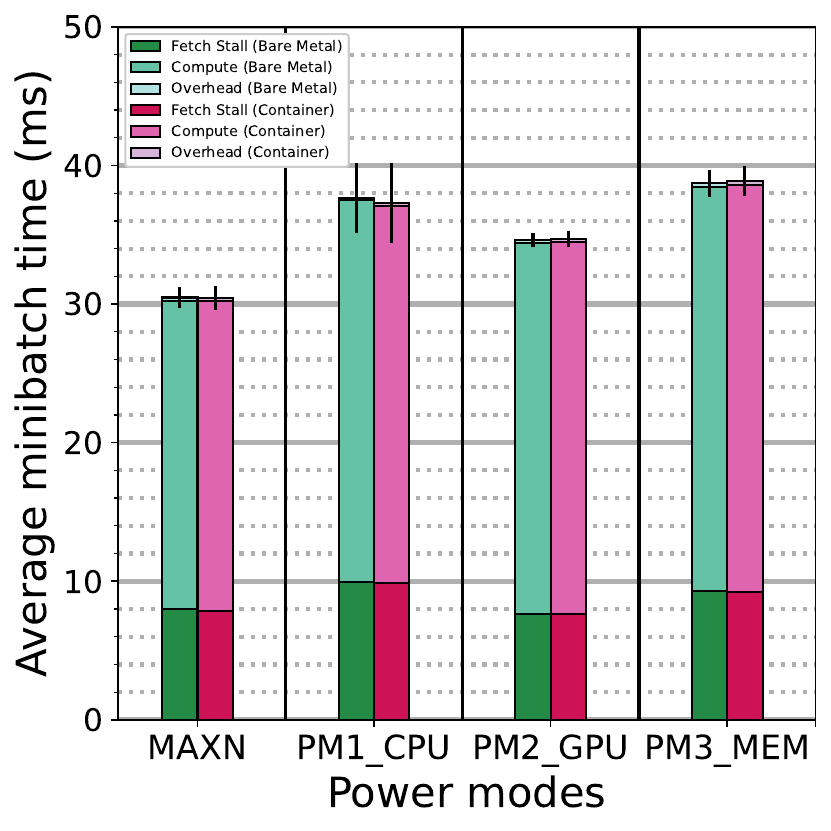}
    \label{fig:infer:powermodes:mobilenet}
  }
\caption{Impact of custom power modes on average minibatch time of DNN training and inference}%
\label{powermodes_train_infer}
%\vspace{-0.125in}
\end{figure*}

% \begin{figure}[t]
%  \vspace{-0.15in}
% \centering%
%     \subfloat{
%   \includegraphics[width=0.45\columnwidth]{plotsv2/splitup-minibatchtime_mobilenetv3_training_eb.pdf}
%     \label{fig:train:gpu_breakup}
%   }
%  \vspace{-0.1in}
% \caption{Compute time break-up for \mobilenet training} %
% \label{train_gpubreakup}
% %\vspace{-0.25in}
% \end{figure}

\subsection{DNN training}
We run each training model on \BM and container and report the average minibatch time, total energy, CPU \& GPU utilization, power and memory usage.
Every training workload is run for at least $3$ epochs or 10 minutes, whichever is longer. We use a minibatch size of $16$ and SGD as the optimizer for all models following standard implementations. Our analysis is presented as a series of takeaways.

%\claim{Lightweight training workloads see a significant increase in compute time when run in a container as opposed to \BM}
\claim{Training workloads do not see an increase in compute time when run in a container as opposed to \BM}
In Fig.~\ref{fig:train:mbtime}, we report average minibatch time as a stacked plot of fetch stall, compute time and logging overhead (minimal, not seen in plots). \delc{As seen from the plot, there is an increase in compute time from \BM to container for all $3$ models. For \lenet, the compute time (2nd stack in the plot) increases from $3.91ms$ to $6.01ms$, an increase of $53.7\%$. \mobilenet sees an increase from $82.12ms$ to $95.48ms$, which is $16.26\%$, and \resnet sees an increase from $50.53ms$ to $54.24ms$, which is $7.34\%$. In terms of FLOPS (Table~\ref{tbl:modeldataset}), \lenet is the most lightweight of the $3$, followed by \mobilenet and then \resnet. As can be seen, the increase in compute time follows the same pattern and is more for lightweight models \lenet and \mobilenet.}\addc{As seen from the plot, there is no increase in minibatch time from \BM to container for all $3$ models. The fetch stall (1st stack) and compute times (2nd stack) also do not show any variation. For instance, for \mobilenet, the average minibatch time is $88.89s$ on \BM and $88.58s$ on container, a difference of $0.35\%$. The difference in minibatch time for \resnet is $0.58\%$. \lenet shows a difference of $2.5\%$, but this is because of small minibatch times and the absolute difference is only $0.14ms$. }
%todo-add numbers

%\claim{This increase in compute time is due to CPU overheads of containerization}
\claim{CPU and GPU resource utilization and power consumption are also comparable across \BM and container for training workloads}
\delc{In order to localize the source of the compute increase, w} We look at violin plots of CPU, GPU utilization and power reported in Fig.~\ref{fig:train:cpuutil},~\ref{fig:train:gpuutil} and~\ref{fig:train:power} respectively. \delc{Looking at \mobilenet, we see that the median GPU utilization falls from $90\%$ on the \BM to $77\%$ on the container, indicating that the GPU is spending time waiting. The lower GPU utilization also causes the median power for \mobilenet to fall slightly from $43.7W$ to $40.8W$ as the GPU is major contributor to power. We also notice a higher median CPU utilization of $23\%$ on the container as opposed to $20.67\%$ on \BM. This leads us to hypothesize that the CPU is the source of the compute overhead. For \resnet, the CPU utilization increase and GPU utilization decrease follow the same trend but are much less pronounced, indicating lesser compute overhead. Since \lenet is a very small model with very low GPU utilization, we do not observe a difference here. Most of the power for \lenet is contributed by the baseload as seen in Fig~\ref{fig:train:power}, and we don't see a difference here too.} \addc{Looking at \mobilenet in Fig~\ref{fig:train:cpuutil}, we see that CPU utilization is $20.67\%$ on \BM and $22.08\%$ on container. \lenet and \resnet follow a similar pattern, with the CPU utilization marginally (1-2\%) higher for container. Similarly, from Fig~\ref{fig:train:gpuutil}, we observe that GPU utilization for \mobilenet are $91\%$ and $91.5\%$ for bare-metal and container respectively. \resnet and \lenet \BM and container have identical GPU utilizations of $20\%$ and $99\%$ respectively. Considering Fig~\ref{fig:train:power}, we see that median power for \mobilenet is $43.21W$ on \BM and $43.61W$ on container. \lenet has a median power of $18.1W$ and $18.5W$ on \BM and container, and \resnet has a median power of $51.51W$ on both. Since the GPU is the major contributor to power, comparable utilizations across \BM and container lead to the same power load.}

%\claim{This increase in compute time is further exacerbated when CPU frequency is lowered using a custom power mode, confirming that the CPU is the source of the overhead}
\claim{Further, we do not see a significant increase in compute time even on custom power modes for training workloads.}
\delc{To verify that the CPU is indeed the source of the overhead, w} We define $3$ custom power modes as listed in Table~\ref{tbl:power} and run the training workloads for these $3$ power modes apart from the default power mode (MAXN) to check if there are any overheads. Each of the power modes lowers one of CPU, GPU and memory frequencies while keeping the others the same as MAXN. 
\delc{\lenet and \mobilenet container runs show a high sensitivity to lowering of CPU frequency more than GPU or memory--i.e. the compute time increases more sharply for container training when CPU frequency is lowered. In going from power mode MAXN to PM1\_CPU, the compute time for \mobilenet on \BM increases from $82.12ms$ to $91ms$, an increase of $10.7\%$. In contrast, for the container, the increase in compute time is from $95.4ms$ to $130ms$, which is a much higher increase of $36.1\%$. We do not see this with other power modes. This proves conclusively that there is a CPU overhead of containerization which affects lightweight models more, and this effect gets worse if run in a power mode with reduced CPU frequency.}
\addc{For \mobilenet, as seen from Fig.~\ref{fig:train:powermodes:mobilenet}, all power modes show similar minibatch times on \BM and container. For \resnet, PM\_MEM shows a very slight ($\approx 1\%$) increase in minibatch time on container. For \lenet, we do see some variation in minibatch times across \BM and container. However, since the absolute minibatch times are very small (less than $10ms$) and show increased variability as indicated by the longer error bars, we do not consider this as a significant difference.}

%\claim{\delc{Why the overhead?}}
\delc{To further narrow down the reason for the CPU overhead, we traced the system calls for \mobilenet training on \BM and container using the Linux \textit{strace} utility. We observed a larger number of cumulative system calls for the container. Since this pointed to a system call related overhead, we ran the container with \textit{seccomp} turned off to observe the effect of Docker's system call filtering. However, we did not observe any change in runtimes from the previous experiments. On a parallel track, we added more fine-grained profiling to the compute phase of the workload, recording times for forward pass, backward pass and the parameter update (\texttt{optimizer.step} in PyTorch). This breakup of the minibatch time is reported in Fig~\ref{train_gpubreakup}. We noticed that among the three compute steps, the parameter update using the SGD optimizer sees the most increase from $2.63ms$ to $17.63ms$, (a $6.7\times$ increase as compared to $11.4\%$ increase for forward and $4.89\%$ increase for backward). We notice that container runs take longer after adding this fine-grained instrumentation and this might be due to asynchronous operations being affected by the \texttt{torch.synchronize} needed for recording the times. This needs further investiagtion. We also performed runs with $0$ DataLoader workers to decouple the CPU overheads of the fetch/pre-process phase from the compute phase, but still observe the same overheads as before.}

%\claim{Energy consumption of the containerized training workload is higher than on \BM, more so for lightweight models}
\claim{Energy consumption of the containerized training workload matches \BM.}
In Fig~\ref{fig:train:energy}, we report the energy consumed by the $3$ models for training as a stacked bar plot. The bottom stack of the bar represents the baseload, the energy it takes for the system to run with no workload. To measure this, we run our logging script for $10$min on \BM with no workload running. Similarly, we spin up a container and run the logging script alone for $10$min. We scale both these energy values by the runtime of the actual workload on the \BM and container respectively. The top stack reports the incremental energy spent for training i.e, energy over and above the baseload needed for model training. 
\delc{Again, we observe that there is an increase in total energy consumption for containerized training as compared to \BM. This increase is around $48.89\%$ for \lenet, $11.11\%$ for \mobilenet, and $6.84\%$ for \resnet, more significant for lightweight models.}

%\claim{This increase in energy for lightweight models is due to running for a longer time at a lower GPU utilization}
\delc{As seen from Fig~\ref{fig:train:mbtime}, there is an increase in average minibatch time. The increase in energy is majorly due to the baseload increase of running for a longer time. For example, for \mobilenet, we see a baseload increase of $30.76\%$ along with an incremental energy increase of $2.94\%$. This indicates that running for longer at a lower GPU utilization is not beneficial i.e, the power benefits obtained by running at a lower GPU utilization are outweighed by the energy consumed by running longer.}\addc{We see some slight variations in the base and incremental energy components, but the difference in total energy is less than $1\%$ for all $3$ models. The total energy difference between container and \BM is $0.58\%$, $0.75\%$ and $0.14\%$ for \lenet, \mobilenet and \resnet respectively.}

\claim{The memory footprint overhead of containerization is minimal for training workloads} %
In Fig~\ref{fig:train:mem}, we report the memory used on \BM v/s container as a violin plot. We also report the base memory usage of the system (no workload) using markers. As expected, we do not observe a large memory overhead for containers. The median memory increases from $4.66$GB to $4.86$GB, an overhead of $0.2$GB for \lenet. Similarly, \mobilenet and \resnet see an increase of $0.35$GB and $0.15$GB respectively. This indicates that multiple containers can be spun up without incurring a significant memory overhead even on resource-constrained edge devices. %

\subsection{DNN inference}

We run each inference model on \BM and container and report the average minibatch time, total energy, CPU \& GPU utilization, power and memory usage. We simulate the effect of images arriving over the network by running a Dataloader iteration before the start of the workload and verify that all images are in memory. Every inference workload is run for at least a few minutes, and the number of minibatches is chosen to ensure this. \resnet is run for $6250$ minibatches, \mobilenet for  $24600$ minibatches, and \lenet for $131250$ minibatches. Minibatch size is set to $16$ for all.

%\claim{Lightweight inference workloads also see an increase in compute time, but the increase is less significant as compared to training}
\claim{Inference workloads also do not see any increase in minibatch time}
In Fig~\ref{fig:infer:mbtime}, we report average minibatch time as a stacked bar plot of fetch stall, compute and overhead times. \delc{We observe that compute time increases by $30.52\%$ for \lenet and $12.93\%$ for \mobilenet. \resnet does not have a noticeable increase in compute time (less than $1\%$). This can be explained by the fact that inference involves only the forward pass, and the parameter update which caused most of the overhead in training is not a part of inference workloads.}\addc{Similar to training, we observe that minibatch time is comparable across bare metal and container for inference workloads too. The difference in minibatch times across \BM and container is $1.1\%$, $0.13\%$ and $0.04\%$ for \lenet,  \mobilenet and \resnet respectively. %Mention numbers
}

We also perform experiments with the custom power modes for \mobilenet and report the average minibatch times in Fig~\ref{fig:infer:powermodes:mobilenet}.

\delc{For the power mode with lowered CPU frequency (PM1\_CPU), we observe that the compute time for the container increases by $33.72\%$ while for the \BM it increases by $24.26\%$ as compared to MAXN. The compute increase for containerized inference is lesser compared to training.}\addc{The performance of bare metal and container is comparable across $3$ out of $4$ power modes. For the power mode with lowered CPU frequency (PM1\_CPU), we observe that bare metal has a minibatch time of $37.65s$ and container $37.26$, a difference of $1\%$. However, we notice higher variability in minibatch times for this power mode as indicated by the longer error bars and this difference in time is within the error bar margin.}

%\claim{Correspondingly, the energy increase is also modest for inference}
\claim{Correspondingly, inference energy is comparable across bare metal and container}
In Fig~\ref{fig:infer:energy}, we report the energy consumed by the $3$ models for inference as a stacked bar plot of baseload and incremental energy. \delc{The increase in total energy is $18.1\%$, $6.63\%$ and $2.68\%$ for \lenet, \mobilenet and \resnet respectively, which is much lower than in training. This is due to the lower overheads of compute time in inference. Again, most of the energy increase comes from the baseload. For instance, \mobilenet sees an increase of $20.03\%$ in the baseload and $0.5\%$ in incremental energy.} \addc{We see slight variations in the base and incremental energy components, but the difference in total energy under $1\%$ for all $3$ models. The total energy in container is higher than \BM by $0.81\%$, $0.86\%$ and $0.72\%$ for \lenet, \mobilenet and \resnet respectively.}

\claim{The memory overhead of containerization for inference is also minimal}
The memory footprint of containerized inference is higher by $0.19$GB, $0.34$GB and $0.17$GB for \lenet, \mobilenet and \resnet respectively as compared to \BM. 

\begin{table}[t]
\centering
\footnotesize
\caption{Power Modes Evaluated on AGX Orin}
\label{tbl:power}
\begin{tabular}{c|rrrr}
\toprule
 \bf{Label} & \bf{CPU Cores} & \bf{CPU MHz} & \bf{GPU MHz} & \bf{RAM MHz} \\
\hline
 \textbf{MAXN}  & {12} & 2201.6 & 1301 & 3200 \\
  
  \textbf{PM1\_CPU}  & {12} & \textbf{1497.6} & 1301 & 3200\\
 
  \textbf{PM2\_GPU}  & {12} & 2201.6 & \textbf{930.75} & 3200 \\
 
  \textbf{PM3\_MEM}  & {12} & 2201.6 & 1301 & \textbf{2133} \\
\bottomrule
\multicolumn{5}{c}{\textit{Cells in bold indicate a value change from the cell in the first row.}}\\
\end{tabular}
% \vspace{-0.2in}
\end{table}

\section{Conclusions and Future Work} %
In this paper, we closely examine and characterize containerized DNN training and inference workloads. \addc{We demonstrate that containerization overheads are minimal for both DNN training and inference workloads. Additionally, we also examine containerization overheads under different power modes and show that workload execution times do not vary.}\delc{We demonstrate that lightweight DNN models incur overheads in compute time and consequently energy when containerized. These overheads show up even more in power modes with lower CPU frequencies. This is relevant to federated learning workloads, which often involve lightweight models.} In future, we plan to \delc{localize the overheads of containerization and also} investigate the effect of containerization when inference and training are run concurrently, as this is an emerging usecase of continuous or lifelong learning. \addc{We also plan to examine the overheads of Firecracker Micro VMs on edge devices as they offer enhanced security.}

\section*{Acknowledgments} We thank members of the DREAM Lab, particularly  Beautlin S and Tuhin Khare for plotting help and Ajay Nayak from CSL Lab for inputs on Docker.

% \vspace{-0.25cm}

\bibliographystyle{IEEEtran}
\bibliography{arxiv}

\end{document}